\documentclass[aps,prb,preprint,superscriptaddress,showkeys,floatfix]{revtex4}

\usepackage{graphicx,color}
\usepackage{multirow,slashbox}

\newcommand{\jwj}[1]{\textcolor{red}{#1}}
\graphicspath{{figs/}}
\begin{document}
\title{A Stillinger-Weber Potential for Single-Layer Black Phosphorus, and the Importance of Cross-Pucker Interactions for Negative Poisson's Ratio and Edge Stress-Induced Bending}
\author{Jin-Wu Jiang}
    \altaffiliation{Corresponding author: jwjiang5918@hotmail.com}
    \affiliation{Shanghai Institute of Applied Mathematics and Mechanics, Shanghai Key Laboratory of Mechanics in Energy Engineering, Shanghai University, Shanghai 200072, People's Republic of China}
\author{Timon Rabczuk}
    \affiliation{Institute of Structural Mechanics, Bauhaus-University Weimar, Marienstr. 15, D-99423 Weimar, Germany}
    \affiliation{School of Civil, Environmental and Architectural Engineering, Korea University, Seoul, South Korea }
\author{Harold S. Park}
    \altaffiliation{Corresponding author: parkhs@bu.edu}
    \affiliation{Department of Mechanical Engineering, Boston University, Boston, Massachusetts 02215, USA}

%\date{22 December 2009}
\date{\today}
\begin{abstract}
The distinguishing structural feature of single-layer black phosphorus is its puckered structure, which leads to many novel physical properties.  In this work, we first present a new parameterization of the Stillinger-Weber potential for single-layer black phosphorus.  In doing so, we reveal the importance of a cross-pucker interaction term in capturing its unique mechanical properties, such as a negative Poisson's ratio.  In particular, we show that the cross-pucker interaction enables the pucker to act as a re-entrant hinge, which expands in the lateral direction when it is stretched in the longitudinal direction. As a consequence, single-layer black phosphorus has a negative Poisson's ratio in the direction perpendicular to the atomic plane.  As an additional demonstration of the impact of the cross-pucker interaction, we show that it is also the key factor that enables capturing the edge stress-induced bending of single-layer black phosphorus that has been reported in {\it ab initio} calculations.
\end{abstract}

%\pacs{68.65.Ac, 62.25.-g}
\maketitle
\pagebreak
\section{Introduction}
Few-layer black phosphorus (BP) is a recent addition to the canon of two-dimensional materials that has been explored as an alternative electronic material to graphene, boron nitride, and the transition metal dichalcogenides for transistor applications\cite{LiL2014,LiuH2014,BuscemaM2014}. This initial excitement surrounding BP is because unlike graphene, BP has a direct bandgap that is layer-dependent.  Furthermore, BP also exhibits a carrier mobility that is larger than MoS$_{2}$\cite{LiuH2014}. The van der Waals effect in bulk BP was discussed by Appalakondaiah et.al.\cite{AppalakondaiahS2012prb} A comprehensive investigation on the strain effect and edge effect on the electronic properties for BP nanoribbons were performed by Guo et. al.\cite{GuoH2014jpcc}  First-principles calculations show that single-layer BP (SLBP) has a band gap around 0.8~{eV}, and that the band gap decreases with increasing thickness\cite{LiuH2014,DuY2010jap}. For SLBP, the band gap can be manipulated via mechanical strain in the direction normal to the BP plane, where a semiconductor-metal transition was observed\cite{RodinAS2014}; while for bilayer BP, the band gap can be affected by the stacking order.\cite{DaiJ2014jpcl}

\begin{figure}[htpb]
  \begin{center}
    \scalebox{0.85}[0.85]{\includegraphics[width=8cm]{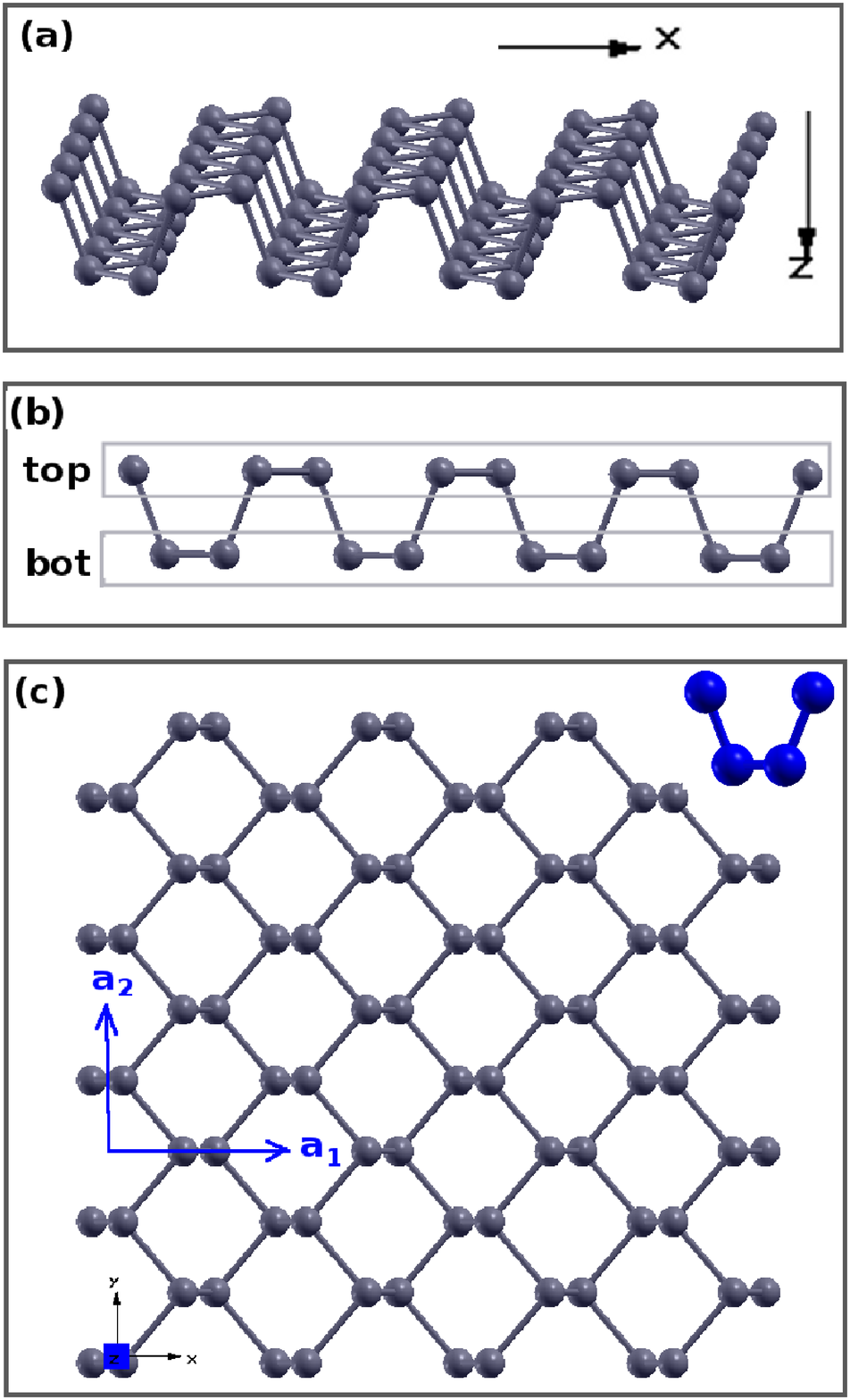}}
  \end{center}
  \caption{(Color online) Configuration for SLBP (4, 5) viewed in different directions. (a) Perspective view. (b) Side view. P atoms are divided into the top and bottom groups. (c) Top view. Inset shows the unit cell with four P atoms in the SLBP. $\vec{a}_{1}$ and $\vec{a}_{2}$ are the two in-plane lattice basis vectors. The x-axis is perpendicular to the pucker. The y-axis is parallel to the pucker. The z-axis is perpendicular to the SLBP plane.}
  \label{fig_cfg}
\end{figure}

\jwj{SLBP has a characteristic puckered structure as shown in Fig.~\ref{fig_cfg}~(a), which leads to two anisotropic in-plane directions. Fig.~\ref{fig_cfg}~(b) shows the side view of the SLBP, which indicates that SLBP can be regarded as the corrugated structure of an atomic plane under external compression. This corrugated structure leads to the anisotropy in the x and y directions.} As a result of this puckered configuration, anisotropy has been found in various properties for SLBP, such as the optical properties\cite{XiaF2014nc,TranV2014prb,LowT2014prb}, the electrical conductance\cite{FeiR2014nl}, and the mechanical properties\cite{AppalakondaiahS2012prb,QiaoJ2014nc,JiangJW2014bpnpr,JiangJW2014bpyoung,QinGarxiv14060261}, including the Poisson's ratio\cite{JiangJW2014bpnpr,QinGarxiv14060261}. \jwj{For instance, in the optical absorption spectra of SLBP, the band edge is 1.55~eV and 3.14~eV in the two in-plane directions, respectively.\cite{XiaF2014nc}  Furthermore, the Poisson's ratio is positive in the x-direction, but negative in the y-direction for SLBP.\cite{JiangJW2014bpnpr}}

However, all of the above reports on the physical properties of SLBP were reported using {\it ab initio} calculation methods.  These methods are well-known to provide relatively accurate results for the physical property of interest, with the downside that the calculations are often very computationally demanding and time-consuming.  Furthermore, due to the complexity of the {\it ab initio} calculations, an intuitive interpretation for the atomic interactions that control the unique physical properties may be difficult to ascertain.

In contrast to {\it ab initio} approaches, empirical interatomic potentials decompose the atomic interactions into distinct, well-defined terms, which typically include bond bending, stretching, and multi-body interactions.  Thus, these potentials can be helpful in understanding structural or mechanical properties that are dominated by a particular type of atomic interaction, while also enabling large-scale molecular dynamics (MD) simulations containing hundreds of thousands or millions of atoms.  Of these interatomic potentials, the Stillinger-Weber (SW) potential is one of the most useful\cite{StillingerFH} as it includes multi-body interactions along with nonlinearity within a relatively simple analytical form that enables high computational efficiency.

In this paper, we first develop a new parameterization of the SW potential for SLBP, while demonstrating the ability of the potential to reproduce the phonon dispersion for SLBP.  More interestingly, we use the SW potential to study the mechanical behavior of SLBP, and reveal an important cross-pucker interaction (CPI) term associated with the puckered structure of SLBP, which describes the interaction strength between two neighboring puckers.  We illustrate that the CPI enables the pucker to act like a re-entrant hinge, which is the key mechanism enabling the recently reported negative Poisson's ratio in the out-of-plane direction\cite{JiangJW2014bpnpr}.  Finally, we demonstrate that the CPI also plays a key role in capturing the edge stress-induced bending of SLBP previously reported in {\it ab initio} calculations\cite{CarvalhoA2014bprb}.

\section{Stillinger-Weber parameterization}

\jwj{Fig.~\ref{fig_cfg} shows the puckered configuration of SLBP. Fig.~\ref{fig_cfg}~(a) shows a perspective view of the pucker in SLBP. The side view shown in Fig.~\ref{fig_cfg}~(b) shows that SLBP can be regarded as the corrugated structure of an atomic plane under external compression. The P atoms in the pucker are categorized into the top and bottom groups in Fig.~\ref{fig_cfg}~(b).} $\vec{a}_{1}$ and $\vec{a}_{2}$ are two basis vectors as shown in Fig.~\ref{fig_cfg}~(c). The size of the SLBP ($n_{1}$, $n_{2}$) is $n_{1}a_{1} \times n_{2}a_{2}$, where $n_{1}$ and $n_{2}$ are the number of unit cells in the x and y-directions. The x-direction is perpendicular to the pucker, the y-direction is parallel to the pucker, and the z-axis is perpendicular to the SLBP plane. The symmetry group for the SLBP is C$_{2s}$ with only one reflection plane symmetric operation at $y=0$.

In 1982, Kaneta, Katayama-Yoshida, and Morita developed a force constant model to describe the lattice dynamics properties of BP\cite{KanetaC1982ssc}. The force constant model provides a reasonable description for BP, which accounts for the covalent bonding between the P atoms.  However, the force constant model contains only linear interactions and thus is not applicable for structural optimization, which searches the energy minimum state.  In contrast, the SW potential includes nonlinear terms and is also suitable for the description of covalent interactions, having originally been developed to model silicon,\cite{StillingerFH} and so we choose to parametrize a SW potential for SLBP.

The SW potential treats the bond bending by a two-body interaction, while the angle bending is described by a three-body interaction. The total potential energy within a system with $N$ atoms is as follows\cite{StillingerFH},
\begin{eqnarray}
\Phi(1,...,N) = \sum_{i<j}V_{2}(i,j) + \sum_{i<j<k}V_{3}(i,j,k).
\end{eqnarray}
The two-body interaction takes following form 
\begin{eqnarray}
V_{2}=\epsilon A\left(B\sigma^{p}r_{ij}^{-p}-\sigma^{q}r_{ij}^{-q}\right)e^{[\sigma\left(r_{ij}-a\sigma\right)^{-1}]},
\label{eq_sw2}
\end{eqnarray}
\jwj{where the exponential function ensures a smooth decay of the potential to zero at the cut-off distance, which is key to conserving energy in MD simulations. The variable $r_{ij}$ is the distance between atoms i and j.  All other parameters in Eq. (\ref{eq_sw2}) will be fitted in the following.}

The three-body interaction is
\begin{eqnarray}
V_{3}=\epsilon\lambda e^{\left[\gamma\sigma\left(r_{ij}-a\sigma\right)^{-1}+\gamma\sigma\left(r_{jk}-a\sigma\right)^{-1}\right]}\left(\cos\theta_{jik}-\cos\theta_{0}\right)^{2},
\label{eq_sw3}
\end{eqnarray}
where $\theta_{0}$ is the equilibrium angle and \jwj{the angle $\theta_{jik}$ is the angle with atom j as the peak.  The other parameters in Eq. (\ref{eq_sw3}) will be determined in the following.}

\begin{figure*}[htpb]
  \begin{center}
    \scalebox{1}[1]{\includegraphics[width=\textwidth]{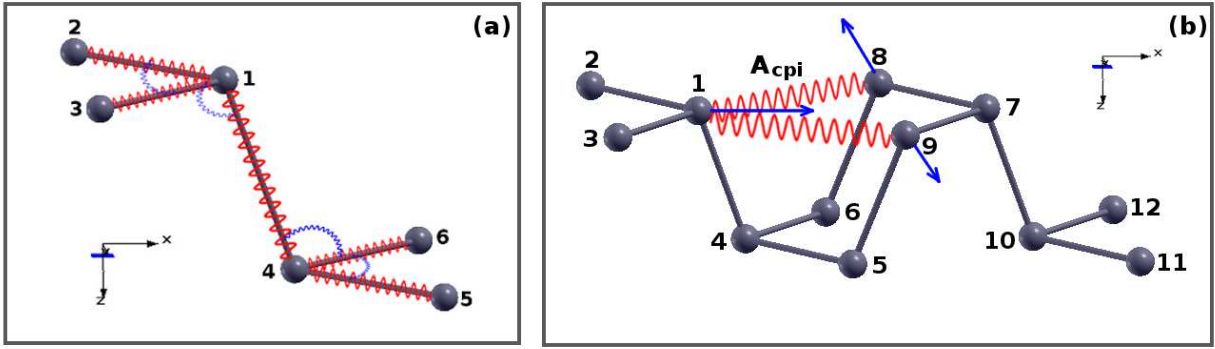}}
  \end{center}
  \caption{(Color online) Atomic structure for SLBP. (a) A single pucker. Each atom P is connected to three FNN atoms. Red springs illustrate the bond bending interactions. Blue springs illustrate the angle bending interactions. (b) Two puckers along the x-axis. The red springs represent the CPI term, with $A_{\rm cpi}$ as the energy parameter. If the structure is stretched (blue arrows on atoms 8 and 9) in the y-direction, the CPI will drag atom 1 (blue arrow) along the x-direction.}
  \label{fig_cfg_hinge}
\end{figure*}

Fig.~\ref{fig_cfg_hinge}~(a) shows the interactions within a single pucker of SLBP. There are five potential terms. (1) $V_{2}(1-2)$ corresponds to the intra-group bond bending between two first-nearest-neighboring (FNN) atoms 1 and 2 (or atoms 4 and 5). (2) $V_{2}(1-4)$ describes the inter-group bond bending between two FNN atoms 1 (top) and 4 (bottom). (3) $V_{3}(2-1-3)$ describes the bending for the intra-group angle $\theta_{213}$ (or $\theta_{546}$). (4) $V_{3}(3-1-4)$ describes the bending for the inter-group angle $\theta_{314}$. (5) $V_{3}(6-4-1)$ describes the bending for the inter-group angle $\theta_{641}$. We have considered the bond stretching and bending energy using the SW potential, while the torsion energy is ignored.

These five potential terms in Fig.~\ref{fig_cfg_hinge}~(a) are between FNN P atoms. Besides these interactions, we find another particularly important two-body interaction as shown in Fig.~\ref{fig_cfg_hinge}~(b), which is the CPI between two second-nearest-neighboring (SNN) atoms 1 and 8 (or 9). \jwj{Atom 1 is from the first pucker (constructed by atoms 1-6), while atom 8 is from a neighboring pucker (constructed by atoms 7-12). The CPI is described by the two-body SW potential as shown in Eq.~(\ref{eq_sw2}). We denote this particular potential term as $V_{2}(1-8)$. The energy parameter for the CPI is $A_{\rm cpi}$.}

\begin{figure}[htpb]
  \begin{center}
    \scalebox{1}[1]{\includegraphics[width=8cm]{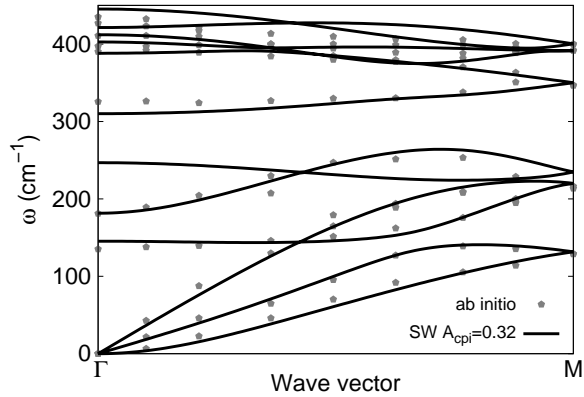}}
  \end{center}
  \caption{(Color online) Phonon dispersion for SLBP along $\Gamma$M from the fitted SW potential is compared to the data from the {\it ab initio} calculation in Ref.~\onlinecite{ZhuZ2014prl}. The CPI has $A_{\rm cpi}=0.32$~{eV}.}
  \label{fig_phonon_fit}
\end{figure}

\begin{table*}
\caption{The two-body (bond bending) SW potential parameters for GULP\cite{gulp}. P atoms are divided into the top (P$_{t}$) and bottom (P$_{b}$) groups. \jwj{The expression is $V_{2}=Ae^{[\rho/\left(r-r_{max}\right)]}\left(B/r^{4}-1\right)$, where $r$ is the distance between two atoms.} Energy parameters have units of eV. Length parameters have units of \AA.}
\label{tab_sw2_gulp}
\begin{tabular*}{\textwidth}{@{\extracolsep{\fill}}|c|c|c|c|c|c|}
\hline
& A & $\rho$ & B & $r_{\rm min}$ & $r_{\rm max}$  \\
\hline
P$_{t}$-P$_{t}$ & 5.5761 & 1.3666 & 12.0666 & 0.0 & 2.79 \\
\hline
P$_{b}$-P$_{b}$ & 5.5761 & 1.3666 & 12.0666 & 0.0 & 2.79 \\
\hline
P$_{t}$-P$_{b}$ & 12.2737 & 2.4633 & 10.9215 & 0.0 & 2.88 \\
\hline
P$_{t}$-P$_{t}$ & 0.32 & 1.3666 & 12.0666 & 2.79 & 3.90 \\
\hline
P$_{b}$-P$_{b}$ & 0.32 & 1.3666 & 12.0666 & 2.79 & 3.90 \\
\hline
\end{tabular*}
\end{table*}

\begin{table*}
\caption{Three-body (angle bending) SW potential parameters for GULP\cite{gulp}. P atoms are divided into the top (P$_{t}$) and bottom (P$_{b}$) groups. The expression is $V_{3}=Ke^{[\rho_{1}/\left(r_{12}-r_{max12}\right)+\rho_{2}/\left(r_{13}-r_{max13}\right)]}\left(\cos\theta-cos\theta_{0}\right)^{2}$, \jwj{where $r_{12}$ is the distance between atoms 1 and 2. The quantity $\theta$ is the angle formed by bonds $r_{12}$ and $r_{13}$}.  Energy parameters have units of eV. Length parameters have units of \AA. Mo-S-S indicates the bending energy for the angle with Mo as the apex.}
\label{tab_sw3_gulp}
\begin{tabular*}{\textwidth}{@{\extracolsep{\fill}}|c|c|c|c|c|c|c|c|c|c|c|}
\hline
& K & $\theta_{0}$ & $\rho_{1}$ & $\rho_{2}$ & $r_{\rm min12}$ & $r_{\rm max12}$ & $r_{\rm min13}$ & $r_{\rm max13}$ & $r_{\rm min23}$ & $r_{\rm max23}$ \\
\hline
P$_{t}$-P$_{t}$-P$_{t}$ & 33.2394 & 99.4656 & 1.0114 & 1.0114 & 0.00 & 2.79 & 0.00 & 2.79 & 0.00 & 3.90   \\
\hline
P$_{b}$-P$_{b}$-P$_{b}$ & 33.2394 & 99.4656 & 1.0114 & 1.0114 & 0.00 & 2.79 & 0.00 & 2.79 & 0.00 & 3.90   \\
\hline
P$_{t}$-P$_{t}$-P$_{b}$ & 162.9172 & 103.307 & 1.5774 & 1.5774 & 0.00 & 2.79 & 0.00 & 2.88 & 0.00 & 3.77   \\
\hline
P$_{b}$-P$_{b}$-P$_{t}$ & 47.5473 & 103.307 & 1.5774 & 1.5774 & 0.00 & 2.79 & 0.00 & 2.88 & 0.00 & 3.77   \\
\hline
\end{tabular*}
\end{table*}

The code GULP\cite{gulp} was used to calculate the fitting parameters for the SW potential, where the SW parameters were fit to {\it ab initio} derived phonon spectrum of SLBP published in Ref.\onlinecite{ZhuZ2014prl}. It is important to ensure that the SW potential can reproduce the phonon spectrum of more accurate, {\it ab initio} methods as the phonon is the most fundamental exciton in solid mechanics. Because the acoustic phonon velocities from the phonon spectrum are closely related to the mechanical properties of the material\cite{LandauLD}, a proper fitting to the phonon spectrum is essential for a good description of the mechanical properties. The number of phonon modes used in the fitting procedure is larger than the number of fitting parameters in the SW potential, so that the resulting parameter set will be unique.

Fig.~\ref{fig_phonon_fit} shows the phonon spectrum calculated by the fitted SW potential and the {\it ab initio} calculation. The fitted SW parameters are shown in Tab.~\ref{tab_sw2_gulp} and Tab.~\ref{tab_sw3_gulp}. \jwj{The calculated acoustic branches agree quite well with the {\it ab initio} calculation, while there are some deviations in the optical branches. The deviation is due to the simplicity of the SW potential implemented in the present work. The SW parameters are fitted to the {\it ab initio} calculation instead of the experiment\cite{YamadaY1984prb}, as the experiment only reported the phonon dispersion in the low-frequency range based on the inelastic neutron scattering method for bulk, and not SLBP.} In the relaxed structure, the two lattice constants are $a_{1}=4.1766$~{\AA} and $a_{2}=3.2197$~{\AA}. The two inequivalent bond lengths are $r_{12}=2.1098$~{\AA} and $r_{14}=2.0349$~{\AA}. The two bond angles are $\theta_{213}=99.466^{\circ}$ and $\theta_{214}=103.307^{\circ}$. These structural parameters are close to the experimental values\cite{BrownA1965ac}.

\section{CPI effect on the phonon spectrum and Young's modulus}

As illustrated in Fig.~\ref{fig_cfg_hinge}~(b), the CPI is a long-range interaction between atoms (1 and 8) from two neighboring puckers. The vector $\vec{r}_{18}$ points from atom 1 to atom 8. This vector has nonzero projection along both the x and y-directions, so the CPI will affect physical or mechanical properties in both directions.

\begin{figure}[htpb]
  \begin{center}
    \scalebox{1}[1]{\includegraphics[width=8cm]{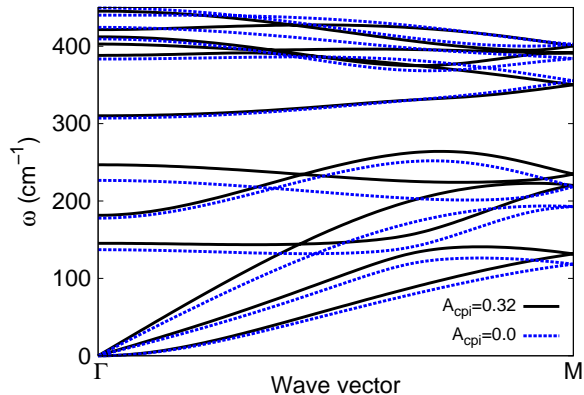}}
  \end{center}
  \caption{(Color online) The effect of CPI on the phonon dispersion of SLBP.}
  \label{fig_phonon_a}
\end{figure}

Fig.~\ref{fig_phonon_a} shows the CPI effect on the phonon spectrum of SLBP. All low-frequency acoustic branches are shifted upward by increasing the CPI. Acoustic phonon modes can propagate information quickly in solids due to their high speeds of sound, i.e. acoustic phonon modes can detect the effects of atoms that are far away.  As a result, the long-range interaction has an important effect on the acoustic branches, and it can increase the frequency of the acoustic branches.

In contrast, the CPI has almost no effect on the optical branches (with frequency above 300~{cm$^{-1}$}) in the phonon spectrum. The optical branches are effectively localized, so they can only detect information for neighboring atoms. As a result, the frequency of the optical branch is determined by the short-range atomic interactions, and is not impacted by the long-range CPI.

\begin{figure}[htpb]
  \begin{center}
    \scalebox{1}[1]{\includegraphics[width=8cm]{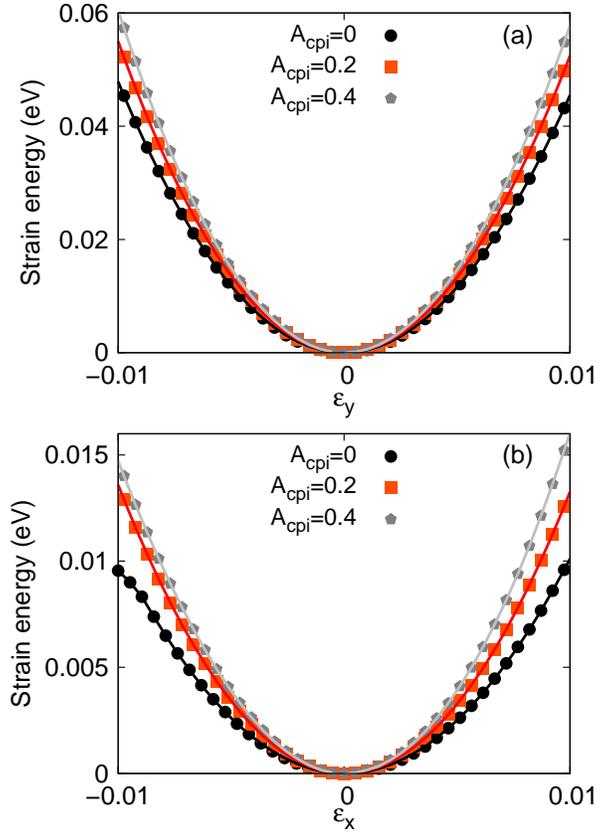}}
  \end{center}
  \caption{(Color online) The effect of CPI on the strain energy of SLBP (4, 5). (a) The SLBP is deformed in the y-direction. (b) The SLBP is deformed in the x-direction.}
  \label{fig_energy}
\end{figure}

\begin{figure}[htpb]
  \begin{center}
    \scalebox{1}[1]{\includegraphics[width=8cm]{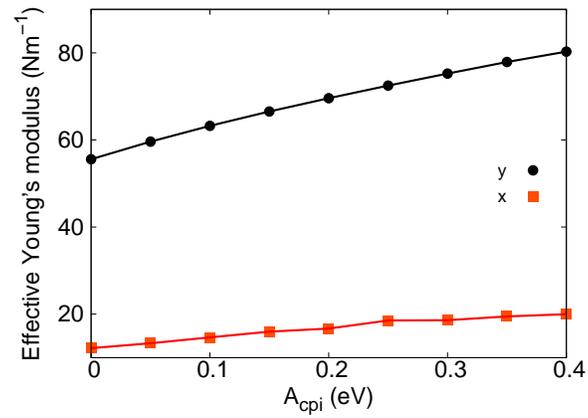}}
  \end{center}
  \caption{(Color online) The effect of CPI on the two-dimensional Young's modulus of SLBP.}
  \label{fig_young_vs_a}
\end{figure}

Because the phonon spectrum is affected by the strength of the CPI, we anticipate that the CPI will impact the Young's modulus of SLBP, because the speed of sound of the acoustic phonon branches are related to the Young's modulus. For a thin plate, the velocities for the two in-plane acoustic phonon branches are $\sqrt{Y/\rho(1-\nu^2)}$, where $Y$ is the Young's modulus, $\nu$ is the Poisson's ratio and $\rho$ is the mass density\cite{LandauLD}.  The frequency for the out-of-plane acoustic branch is proportional to $\sqrt{Yh^2/12\rho(1-\nu^2)}$, where $h$ is the thickness of the plate.

Fig.~\ref{fig_energy} shows the effect of CPI on the strain energy of SLBP (4,5), where periodic boundary conditions were applied in the in-plane, i.e. x and y-directions, while free boundary conditions were applied in the out-of-plane (z)-direction.  For Fig.~\ref{fig_energy}~(a), SLBP is deformed uniaxially in the y-direction with strain $\epsilon_{y}$, while the structure is allowed to be fully relaxed in the other two directions. Fig.~\ref{fig_energy}~(b) is for SLBP deformed uniaxially in the x-direction with strain $\epsilon_{x}$. In both panels, the strain energy increases with increasing CPI.

\jwj{The effective two-dimensional Young's modulus (Y) can be extracted from the strain energy surface density (E) through $E=(1/2)Y\epsilon^2$, where $\epsilon$ is the strain and the strain energy surface density is the strain energy per area.} Fig.~\ref{fig_young_vs_a} shows the effective Young's modulus for SLBP.  The CPI increases the Young's modulus in both x and y-directions, while more strongly impacting the Young's modulus in the y-direction. The Young's modulus is highly anisotropic, and its value in the y-direction is larger than the x-direction by a factor of about four.\jwj{We note that similar anisotropy in the Young's modulus has also been obtained in several previous studies,\cite{AppalakondaiahS2012prb,QiaoJ2014nc,JiangJW2014bpnpr,JiangJW2014bpyoung,QinGarxiv14060261} though the obtained values show variability between the different studies. The difference is probably due to different computational methods and potentials that have been used in different studies. For instance, the Young's modulus from the {\it ab initio} calculations for the x and y-directions is 28.9~{Nm$^{-1}$} and 101.6~{Nm$^{-1}$} in Ref.\onlinecite{QiaoJ2014nc}, or 21.9~{Nm$^{-1}$} and 56.3~{Nm$^{-1}$} in Ref.\onlinecite{JiangJW2014bpyoung}. In present work, the Young's modulus is 19.5~{Nm$^{-1}$} in the x-direction and 78.0~{Nm$^{-1}$} in the y-direction for $A_{\rm cpi}=0.32$~{eV}, which is within the range of the previously reported {\it ab initio} values.}

\section{The effect of CPI on negative Poisson's ratio}

It can be seen from Fig.~\ref{fig_cfg_hinge} that the major function of the CPI is to connect two neighboring puckers in SLBP.  A direct result of this cross-pucker interaction is that the CPI will provide a dragging force (blue arrow) on atom 1 in the x-direction when the structure is stretched (blue arrows) in the y-direction. Hence, the structure will contract more in the x-direction when stretched in the y-direction. As a result, the bond 1-4 becomes more closely aligned with the vertical (z)-direction. As a consequence, the thickness (that is, the projection of bond length $r_{14}$ to the z-direction) is increased, and as a result SLBP expands in the z-direction when stretched in the y-direction, leading to a negative Poisson's ratio in this direction.

\begin{figure}[htpb]
  \begin{center}
    \scalebox{1}[1]{\includegraphics[width=8cm]{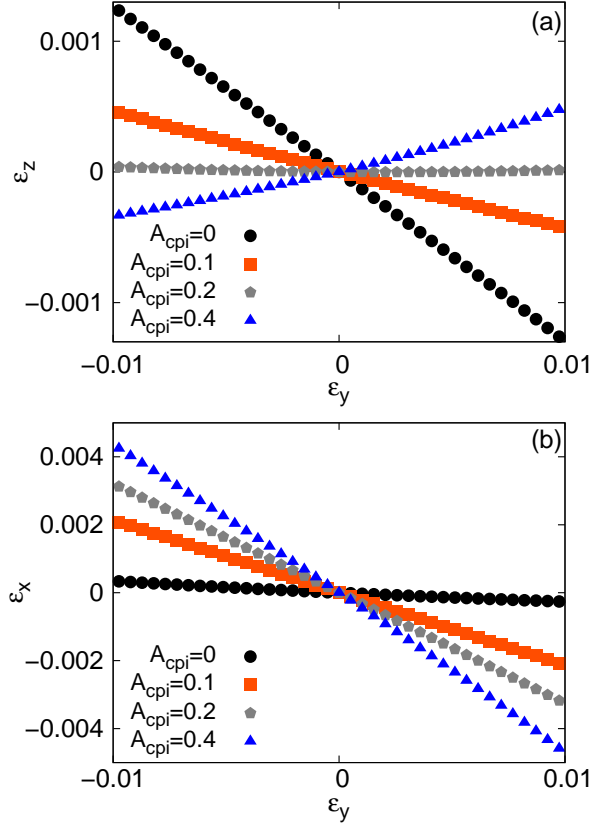}}
  \end{center}
  \caption{(Color online) The effect of CPI on the strain-strain relation for SLBP (4,5) when the SLBP is deformed in the y-direction. (a). The CPI has a strong effect on $\epsilon_{z}$, where the slope of the curve changes sign for $A_{\rm cpi}>0.21$. (b) The effect of CPI on $\epsilon_{x}$.}
  \label{fig_strain_vs_y}
\end{figure}

\begin{figure}[htpb]
  \begin{center}
    \scalebox{1}[1]{\includegraphics[width=8cm]{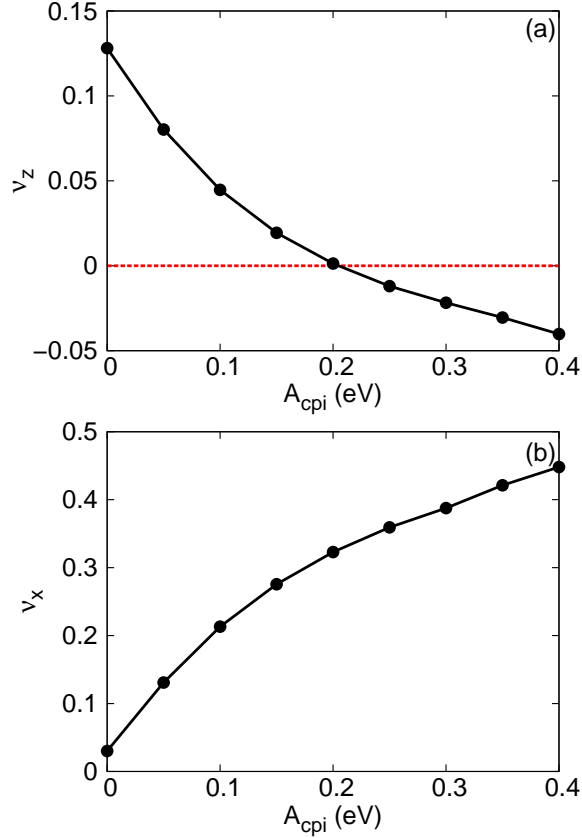}}
  \end{center}
  \caption{(Color online) The effect of CPI on the Poisson's ratio of SLBP (4,5) when SLBP is deformed in the y-direction. (a) The Poisson's ratio in the z-direction ($\nu_{z}$) is decreased by increasing the CPI. The red dashed horizontal line displays $\nu_{z}=0$.  It can be seen that the negative Poisson's ratio phenomenon occurs in the z-direction for $A_{\rm cpi}>0.21$. (b) The Poisson's ratio in the x-direction ($\nu_{x}$) is increased by increasing CPI.}
  \label{fig_poisson_vs_ay}
\end{figure}

\begin{figure}[htpb]
  \begin{center}
    \scalebox{1}[1]{\includegraphics[width=8cm]{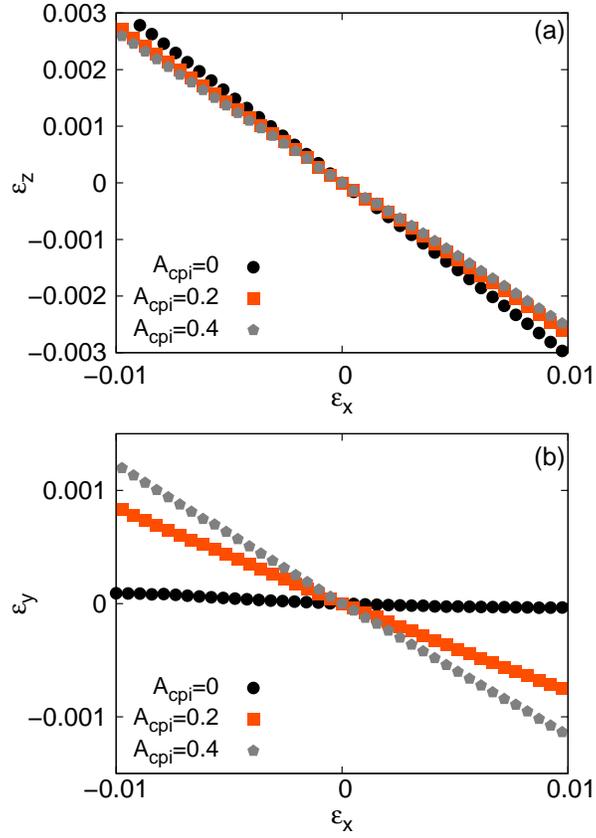}}
  \end{center}
  \caption{(Color online) The effect of CPI on the strain-strain relation of SLBP (4,5) when SLBP is deformed in the x-direction.}
  \label{fig_strain_vs_x}
\end{figure}

\begin{figure}[htpb]
  \begin{center}
    \scalebox{1}[1]{\includegraphics[width=8cm]{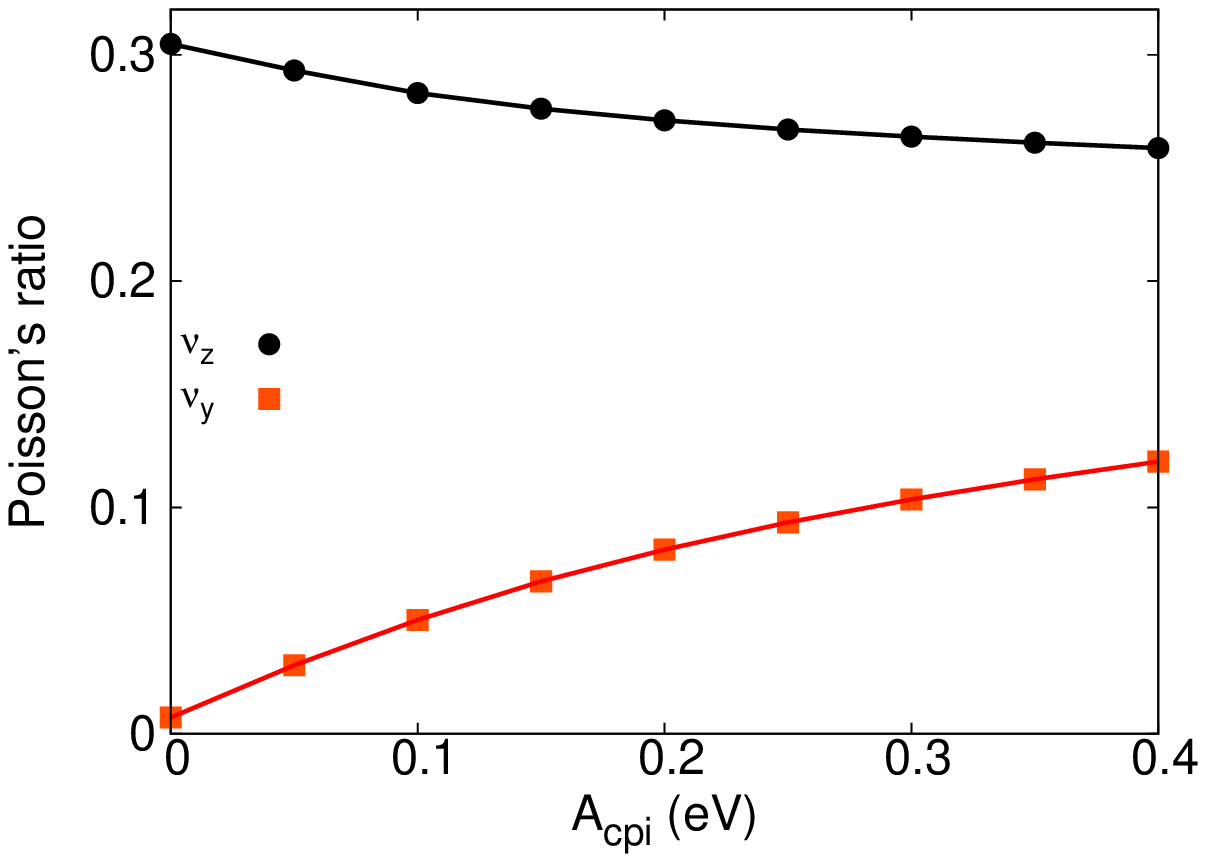}}
  \end{center}
  \caption{(Color online) The effect of CPI on the Poisson's ratio of SLBP (4,5) when SLBP is deformed in the x-direction.}
  \label{fig_poisson_vs_ax}
\end{figure}

Fig.~\ref{fig_strain_vs_y} shows the effect of CPI on the strain in SLBP (4,5) when SLBP is deformed in the y-direction. The Poisson effect leads to the deformation of the structure in the other two directions. The z-direction strain ($\epsilon_{z}$) is evaluated through the change in distance between the top and bottom atoms as previously illustrated in Fig.~\ref{fig_cfg}~(b). Fig.~\ref{fig_strain_vs_y}~(a) shows that the slope of the strain curve changes sign if the CPI is stronger than 0.21~{eV}. \jwj{Accordingly, Fig.~\ref{fig_poisson_vs_ay} shows that the z-direction Poisson's ratio (${\nu_{z}}$) decreases monotonically with increasing CPI, where the z-direction Poisson's ratio is calculated by $\nu_{z}=-\frac{\epsilon_{z}}{\epsilon_{y}}$.} If the CPI is small, $A_{\rm cpi}<0.21$~{eV}, there is no negative Poisson's ratio phenomenon. For $A_{\rm cpi}>0.21$~{eV}, the negative Poisson's ratio phenomenon occurs. This curve explicitly displays the direct correlation between the CPI and the negative Poisson's ratio phenomenon. 

\jwj{As the CPI term has a dominant effect on the negative Poisson's ratio for the SLBP, we determine the value of $A_{\rm cpi}$ by fitting it to the negative Poisson's ratio value. A recent {\it ab initio} calculation obtained $\nu_{z}=-0.027$\cite{JiangJW2014bpnpr}. From the curve in Fig.~\ref{fig_poisson_vs_ay}~(a), we can conclude that the corresponding CPI strength should be $A_{\rm cpi}\approx 0.32$~{eV} in order to match that {\it ab initio} calculation. We stress that the CPI term is important in the sense that it has a dominant effect on the negative Poisson's ratio phenomenon.  We investigated all other SW potential parameters and their effects on the negative Poisson's ratio phenomenon prior to the CPI term.  We found that the phonon dispersion and other properties for SLBP can be reasonably described by the SW potential even without the CPI potential term. However, we found that the Poisson's ratio is always positive if the CPI term is not considered. This is because the CPI term physically represents the origin for the hinge mechanism of the puckered unit cell in the SLBP.}

It is now clear that the CPI is the key atomic interaction that enables the re-entrant functionality of the pucker in SLBP, i.e. it enables the pucker to expand in the z-direction when it is stretched in the y-direction. This type of re-entrant motion is the fundamental mechanism for the negative Poisson's ratio in z-direction in the SLBP.\cite{JiangJW2014bpnpr}

Fig.~\ref{fig_strain_vs_y}~(b) shows the effect of CPI on the Poisson induced strain in the x-direction for SLBP (4,5) that is deformed in the y-direction. The corresponding Poisson's ratio is displayed in Fig.~\ref{fig_poisson_vs_ay}~(b), where it is shown that increasing the CPI enhances $\nu_{x}$ considerably. A large value of $\nu_{x}$ is important for the negative Poisson's ratio in the z-direction, because for larger $\nu_{x}$, SLBP will exhibit larger contraction in the x-direction when it is stretched in the y-direction. Such strong contraction helps to make the bond 1-4 more closely aligned with the z-direction, leading to a more strongly negative Poisson's ratio.

Fig.~\ref{fig_strain_vs_x} shows the effect of CPI on the Poisson induced strains in SLBP (4,5) that is deformed in the x-direction. Fig.~\ref{fig_poisson_vs_ax} displays the corresponding Poisson's ratio value. In this situation, the z-direction Poisson's ratio is reduced by increasing CPI, while the y-direction Poisson's ratio is increased.

\section{CPI and edge stress-induced bending}

In the above, we have established the importance of the CPI effect on SLBP with periodic boundary conditions in the in-plane, or x and y-directions.  However, using periodic boundary conditions removes the effects of free edges, which are known to cause dramatic changes in various physical properties for two-dimensional materials\cite{CastroNAH}.  With regards to mechanical behavior, edge stresses resulting from undercoordinated edge atoms in graphene were shown to lead to warping of monolayer graphene\cite{ShenoyVB,LuQ2010prb}.  However, in contrast to graphene, {\it ab initio} calculations showed that edge stresses in SLBP have been shown to induce bending rather than warping\cite{CarvalhoA2014bprb}.  As the effects of edge stresses on the mechanical behavior of SLBP has not been elucidated extensively, we systematically investigate the effects of edge stresses on the equilibrium configurations of SLBP.

\begin{figure*}[htpb]
  \begin{center}
    \scalebox{0.8}[0.8]{\includegraphics[width=\textwidth]{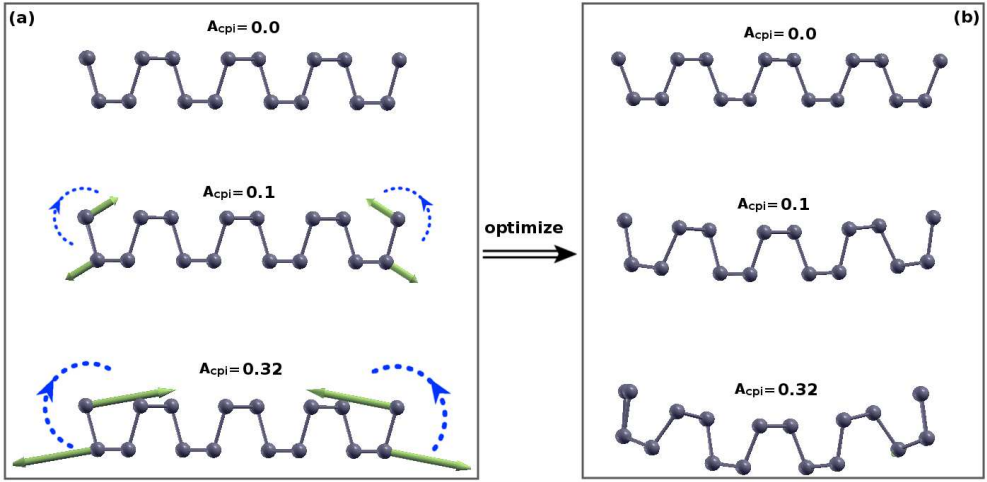}}
  \end{center}
  \caption{(Color online) The structure of SLBP (4,5) with free boundary conditions in the x-direction. (a) Structure before relaxation. Figure sequence from top to bottom corresponds to increasing CPI strength. The arrows on the left and right ends illustrate the edge force. The length of the arrow is proportional to the magnitude of the edge force. The dashed vortexes illustrate the driving force for bending induced by the edge force. (b) Relaxed structure.  The edge force is relaxed by the bending of SLBP during the structural optimization.  Furthermore, increased bending curvature is observed with increasing CPI.  Pronounced bending is observed for $A_{\rm cpi}=0.32$.}
  \label{fig_cfg_fbcx}
\end{figure*}

\begin{figure}[htpb]
  \begin{center}
    \scalebox{1}[1]{\includegraphics[width=8cm]{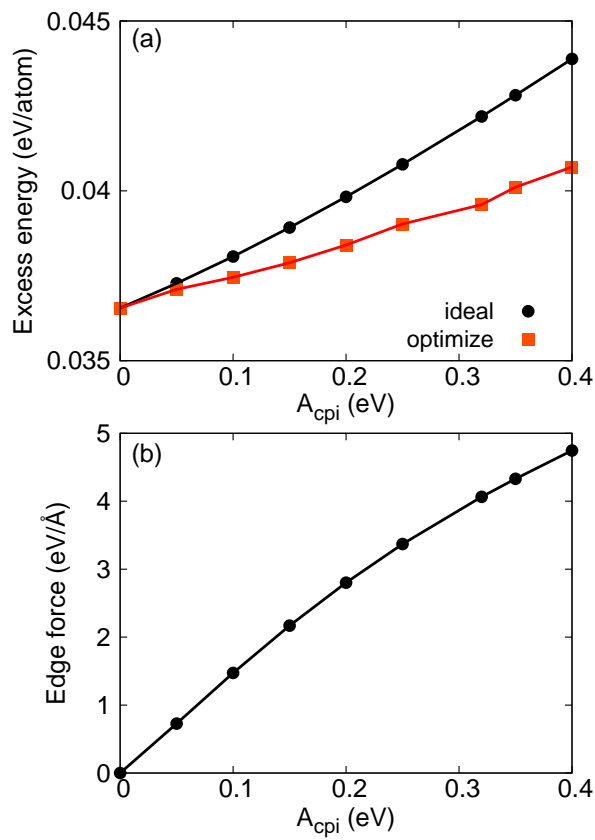}}
  \end{center}
  \caption{(Color online) Edge effect for SLBP (4,5) with free boundary conditions in the x-direction. (a) Excess energy. (b) Edge stress. The solid line is drawn as a guide to the eye.}
  \label{fig_edge_force_a}
\end{figure}

\begin{figure}[htpb]
  \begin{center}
    \scalebox{1}[1]{\includegraphics[width=8cm]{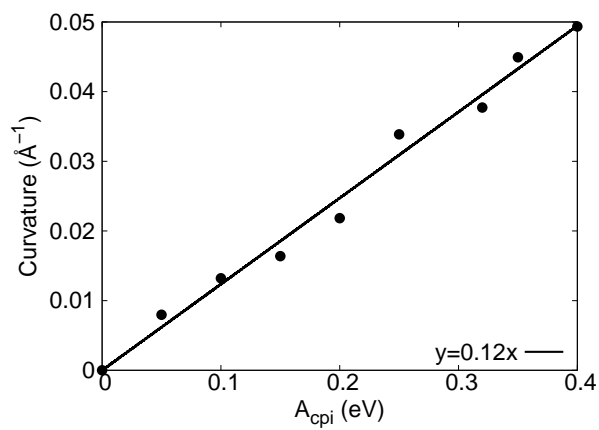}}
  \end{center}
  \caption{(Color online) Edge stress-induced bending curvature for SLBP (4,5) with free boundary conditions in the x-direction. The curvature increases nearly linearly with increasing $A_{\rm cpi}$, where the solid line is drawn as a guide to the eye.}
  \label{fig_fbcx_curvature_n4}
\end{figure}

\begin{figure*}[htpb]
  \begin{center}
    \scalebox{0.8}[0.8]{\includegraphics[width=\textwidth]{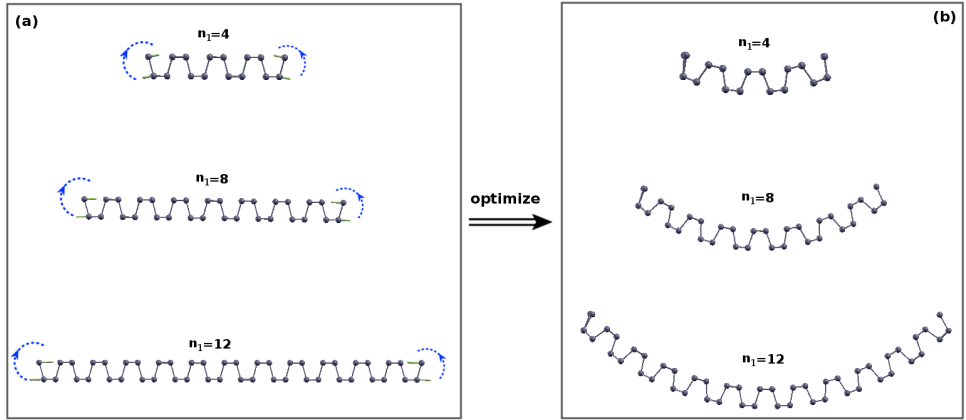}}
  \end{center}
  \caption{(Color online) Edge stress-induced bending for SLBP ($n_{1}$, 5) with free boundary conditions in the x-direction. (a) Structure before relaxation. The CPI has $A_{\rm cpi}=0.32$, and the arrows on the left and right ends illustrate the direction of the edge stress. The dashed vortexes illustrate the driving force for bending induced by the edge stress. (b) Relaxed structure.  SLBP relieves the edge stress by bending during the structural optimization.}
  \label{fig_cfg_fbcx_size}
\end{figure*}

\begin{figure}[htpb]
  \begin{center}
    \scalebox{1}[1]{\includegraphics[width=8cm]{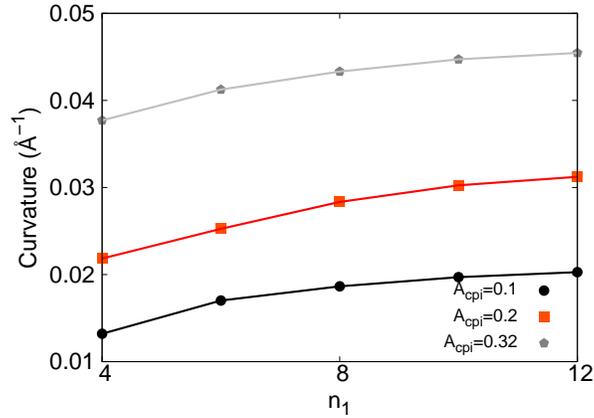}}
  \end{center}
  \caption{(Color online) Edge stress-induced bending curvature versus SLBP length. The bending curvature increases for increasing SLBP length.}
  \label{fig_fbcx_curvature_n}
\end{figure}

\begin{figure*}[htpb]
  \begin{center}
    \scalebox{0.8}[0.8]{\includegraphics[width=\textwidth]{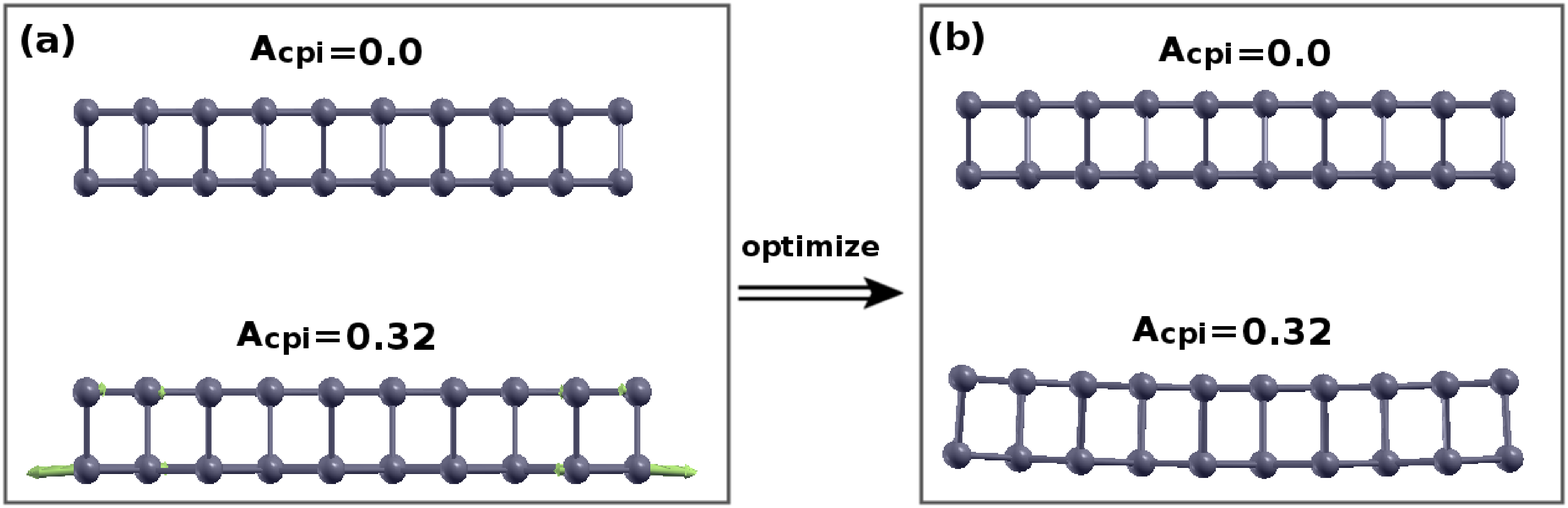}}
  \end{center}
  \caption{(Color online) The relaxed structure for SLBP (4,5) with free boundary conditions in the y-direction. (a) Structure before relaxation. The arrows on the left and right ends illustrate the direction of the edge stress.  No driving force for bending is induced. (b) Relaxed structure. The SLBP is weakly bent along the y-direction due to the edge stress.}
  \label{fig_cfg_fbcy}
\end{figure*}

We first consider SLBP (4,5) that has free edges in the x-direction as shown in Fig.~\ref{fig_cfg_fbcx}, where the left and right ends are free edges. Fig.~\ref{fig_cfg_fbcx}~(a) shows the ideal structure before relaxation (i.e. conjugate gradient energy minimization), while Fig.~\ref{fig_cfg_fbcx}~(b) shows the relaxed structure. SLBP exhibits no edge stress-induced bending without CPI.  However, with increasing CPI the edge stress-induced bending becomes more obvious, with distinctive bending observed for $A_{\rm cpi}=0.32$~{eV}. Similar edge stress-induced bending has also been reported in a recent {\it ab initio} calculation\cite{CarvalhoA2014bprb}.

For graphene, the edge stresses induce warping at the free edges, and because graphene is such a thin plate (one atom thick), the edge stress-induced warping can be localized at the free edge, and decays quickly from the edge into the bulk. However, in contrast to graphene, SLBP actually has two intrinsic atomic layers (top and bottom), so its bending stiffness is larger than graphene.  Because of this, it is harder for SLBP to accommodate the edge stress via local structure reconstruction, so a global shape reconstruction is preferred instead of local warping. Furthermore, Fig.~\ref{fig_cfg_fbcx}~(a) shows that the inward forces compress the top atoms, while the outward forces stretch the bottom atoms, which leads to the driving force for bending shown by the dashed blue vortexes.  This driving force for bending enables the global bending exhibited by SLBP due to the edge stress. \jwj{It should be noted that this driving force induces pure bending on the SLBP, so the torsional potential that has been ignored in this work has no effect on this edge induced bending phenomenon.}

The excess energy and the stress (which is also called the edge force for two-dimensional systems due to having the same units of eV/\AA~\cite{LuQ2010prb}) induced by the free edges are shown in Fig.~\ref{fig_edge_force_a}~(a). The excess energy is defined to be the energy difference per atom between the SLBP with free boundary conditions and SLBP with periodic boundary conditions. The energy per atom is increased after the appearance of free edges. We have shown the excess energy for the SLBP with free edges before and after structural optimization. The excess energy is lowered by structural optimization which induces a global bending on the SLBP.  Fig.~\ref{fig_edge_force_a}~(b) shows the maximum amplitude of the edge force, which increases monotonically with increasing CPI. The curvature for the edge stress-induced bending is displayed in Fig.~\ref{fig_fbcx_curvature_n4}.  As can be seen, the bending curvature increases almost linearly with increasing CPI, as the edge force is larger for larger CPI. These results show that the CPI is the origin for the edge induced bending in the SLBP.

Fig.~\ref{fig_cfg_fbcx_size} shows the relaxed structure for SLBP ($n_{1}$, 5) with different lengths, which all have free edges in the x-direction.  All SLBPs have the same free edge, as they have the same width in the y-direction, and so the edge force is the same for this set of SLBP geometries.  As a result, we can consider the classical Euler bending model, which states that longer beams are easier to bend under the same external force\cite{TimoshenkoS1987}. Consequently, we obtain larger bending curvature for longer SLBP as shown in Fig.~\ref{fig_fbcx_curvature_n}.  

Fig.~\ref{fig_cfg_fbcy} shows the relaxed structure for SLBP (4,5) with free boundary conditions applied in the y-direction.  Fig.~\ref{fig_cfg_fbcy}~(a) shows the structure for the SLBP before optimization, where edge forces are illustrated by arrows.  Fig.~\ref{fig_cfg_fbcy}~(b) shows the relaxed structure for the SLBP. It shows that the SLBP will be slightly bent along the y-direction for stronger CPI with $A_{\rm cpi}=0.32$~{eV}, but that edge stress-induced bending is very weak. The weak bending in the y-direction is because two neighbors are cut off for atoms at the free edge in the x-direction, while only one neighbor is lost for edge atoms in the y-direction, and so the edge effect in the y-direction is weaker. Furthermore, Fig.~\ref{fig_cfg_fbcy}~(a) shows that the outward edge force is on the bottom atoms, while the inward forces on the top atoms are neglectable, so there is almost no bending momentum at the free edges in the y-direction. Hence, the bending in this situation is very weak.

\section{Conclusion}
In conclusion, we have parameterized a Stillinger-Weber potential for the atomic interactions in single-layer black phosphorus.  In doing so, we have revealed the importance of the cross pucker interaction in capturing certain key phenomena that result from the puckered configuration.  In particular, the cross pucker interaction provides the re-entrant mechanism for the pucker in single layer black phosphorus, leading to a negative Poisson's ratio in the out-of-plane direction.  We also demonstrated, as a second example, the importance of the cross pucker interaction in capturing the edge stress-induced bending of single layer black phosphorus. Overall, a cross pucker interaction value of about $A_{\rm cpi}=0.32$ eV enables capturing the negative Poisson's ratio value previously reported in {\it ab initio} calculations.

\textbf{Acknowledgements} We thank Jing-Tao L$\rm \ddot{u}$ at HUST for useful communications on the edge induced bending phenomenon. The work is supported by the Recruitment Program of Global Youth Experts of China and the start-up funding from Shanghai University. HSP acknowledges the support of the Mechanical Engineering department at Boston University.

%\bibliographystyle{aipnum4-1}
%\bibliographystyle{model3-num-names}
%\bibliography{biball}
%\bibliography{/home/JiangJinWu/Documents/papers/mypapers/latex/biball}
\end{document}